\documentclass[twocolumn]{aastex631}

\usepackage{soul}
\usepackage{bm}
\usepackage{amsmath}
\usepackage{tikz}
\usetikzlibrary{shapes.geometric, arrows, positioning, fit, backgrounds}

\tikzset{
    block_input/.style = {rectangle, rounded corners, minimum width=3cm, minimum height=1cm, text centered, draw=blue!70, fill=blue!5},
    block_scaling/.style = {rectangle, minimum width=3.5cm, minimum height=1cm, text centered, draw=orange!70, fill=orange!5},
    block_sim/.style = {rectangle, minimum width=3.5cm, minimum height=1cm, text centered, draw=green!60!black, fill=green!5},
    block_final/.style = {rectangle, double, minimum width=3.5cm, minimum height=1cm, text centered, draw=purple!70, fill=purple!5},
    arrow/.style = {thick, ->, >=stealth}
}
\begin{document}
\newcommand{\fhc}[1]{{\color{blue}{#1}}}
\newcommand{\wjc}[1]{{\color{red}{#1}}}
\newcommand{\mbh}{\ifmmode {M_{\bullet}} \else $M_{\bullet}$\fi}
\newcommand{\mvp}{\ifmmode {M_{\rm VP}} \else $M_{\rm VP}$\fi}
\newcommand{\rblr}{\ifmmode {R_{\rm BLR}} \else $R_{\rm BLR}$\fi}
\newcommand{\kms}{$\rm km\,s^{-1}$}
\newcommand{\feii}{\ensuremath{\mathrm{Fe~\textsc{ii}}}}
\newcommand{\hb}{\ensuremath{\mathrm{H}\beta}}

\title{The Impact of Elliptical Broad-Line Regions on Reverberation-Based Black Hole Mass Estimates}

\correspondingauthor{Hai-Cheng Feng}
\email{hcfeng@ynao.ac.cn}
\correspondingauthor{Xinwu Cao}
\email{xwcao@zju.edu.cn}

\author[0000-0002-2581-8154]{Jiancheng Wu}
\affiliation{Institute for Astronomy, School of Physics, Zhejiang University, 866 Yuhangtang Road, Hangzhou 310058, People’s Republic of China}

\author[0000-0002-1530-2680]{Hai-Cheng Feng}
\affiliation{Yunnan Observatories, Chinese Academy of Sciences, Kunming 650216, Yunnan, People’s Republic of China}

\author[0000-0003-4773-4987]{Qingwen Wu}
\affiliation{Department of Astronomy, School of Physics, Huazhong University of Science and Technology, Luoyu Road 1037, Wuhan, China}

\author[0000-0002-2355-3498]{Xinwu Cao}
\affiliation{Institute for Astronomy, School of Physics, Zhejiang University, 866 Yuhangtang Road, Hangzhou 310058, People’s Republic of China}

\begin{abstract}
The virial factor $f$ is critical for accurate supermassive black hole (SMBH) mass measurements using reverberation mapping (RM) and the radius--luminosity ($R$--$L$) relation, yet its value remains highly uncertain. While traditional models assume axisymmetric broad-line region (BLR) geometries, growing evidence suggests that BLRs may possess more complex, asymmetric structures. We systematically investigate the impact of elliptical-disk BLR geometries on SMBH mass determinations through comprehensive numerical simulations. By computing emission-line profiles, emissivity-weighted time lags, and the corresponding virial factor $f$ over a wide range of eccentricities, orientations, and inclinations, we find that even in purely virialized systems, geometric effects alone can cause $f$ to vary by more than an order of magnitude and can mimic observational signatures typically attributed to radiation pressure. Additionally, local broadening introduces further systematic uncertainties in velocity width measurements, biasing $f$ by up to a factor of $\sim$3. Asymmetric BLR configurations also induce a scatter of $\sim$0.18 dex in the $R$--$L$ relation due to projection effects, comparable to the intrinsic scatter observed in RM studies. These results challenge the conventional attribution of RM uncertainties to non-virial motions or radiation pressure, and instead highlight the fundamental role of BLR geometry in SMBH mass measurements.

\end{abstract}


\section{Introduction} \label{sec:1}
Active galactic nuclei (AGNs) are powered by accretion of matter onto supermassive black holes (SMBHs), where the accretion disk emits a strong continuum spectrum ranging from ultraviolet to near-infrared wavelengths \citep{Shakura1973}. This intense radiation photoionizes the gas surrounding the accretion disk, producing prominent emission lines. These ionized gas clouds are generally believed to be gravitationally bound to the central SMBH and to move at velocities of thousands of \kms. The Doppler broadening effect results in the characteristic broad emission lines of Type I AGNs \citep{Antonucci1993}. This region, known as the broad-line region (BLR), encodes dynamical information about gas within the gravitational potential of the central black hole. Therefore, understanding the distribution and velocity field of BLR gas is essential for the accurate measurement of SMBH masses.

For a virialized BLR, the characteristic radius (\rblr) and the emission-line velocity width ($v$) are expected to follow the virial relation $\rblr \propto v^{-2}$ \citep{Peterson1999, Peterson2000, Kollatschny2003}. In this idealized case, one can define a scale-invariant estimate of the black hole mass, the so-called virial product:
\begin{equation} \label{eq:1}
\mvp = \frac{\rblr v^2}{G},
\end{equation}
where $G$ is the gravitational constant. However, in realistic BLRs, the observed line width, generally characterized by full width at half maximum (FWHM) or line dispersion ($\sigma$), represents a line-of-sight projection of the three-dimensional velocity field, which is sensitive to the inclination and geometry of the BLR \citep{Peterson2001}. Moreover, gravitational forces and radiation pressure can also lead to non-virial motions, such as inflows and outflows \citep{Denney2009}. As a result, the virial product may systematically deviate from the true black hole mass $\mbh$. Therefore, a dimensionless virial factor $f$ is usually introduced to account for these effects, with the black hole mass estimated as $\mbh = f \cdot \mvp$ \citep[e.g.,][]{Bentz2009, Du2018, Feng2021a, Malik2023}.

Direct spatially resolved measurements of the BLR are extremely challenging due to the small angular scales involved. To date, only GRAVITY interferometry has provided limited spatial constraints on the BLR in a few AGNs \citep{Gravity2018, gravity2020, Gravity2021, GRAVITY2024, Abuter2024}. A more widely used method is the reverberation mapping (RM) technique, which exploits the time delay between continuum variations and the corresponding emission-line response to infer the BLR size \citep{Blandford1982, Peterson1993}. This lag ($\tau$) reflects the light-travel time of ionizing photons across the BLR, and the BLR radius is estimated as $R_{\tau} = c\tau$, where $c$ is the speed of light.

Through extensive RM campaigns over the past decades, an empirical correlation has emerged between the BLR size and AGN luminosity, known as the $R$--$L$ relation \citep[e.g.,][]{Kaspi2000, Bentz2013, Grier2017, Du2019, Dalla2020, McDougall2025}. This relation allows the BLR size to be estimated from a single-epoch spectrum, making it a fundamental tool for black hole mass estimation across large AGN samples.

In most RM campaigns, the time lag can be measured with a precision better than 50\% \citep[e.g.,][]{Bentz2010, Shen2016, Feng2021a, Li2022a, Cho2023, Pandey2024}, and often even better than 20\% \citep[e.g.,][]{Bentz2008, Oknyansky2021, Feng2021b, Feng2025, Yao2024}. Consequently, the dominant source of uncertainty in RM-based black hole mass estimates is widely attributed to the $f$ factor \citep[e.g.,][]{Park2012, Grier2013, Williams2018}. In the case of the $R-L$ relation, recent studies have shown that its intrinsic scatter can reach up to 0.3 dex, thereby introducing additional uncertainty in the inferred BLR size \citep{Du2018, Wang2024, Li2026}. Traditional RM commonly models the BLR as a centrally symmetric geometry, such as a disk or a spherical distribution of ionized clouds. Under this assumption, the uncertainty in the $f$ factor is generally considered to arise from the inclination angle of the BLR and the influence of radiation pressure \citep{Collin2006, Pancoast2014, Yong2016, Liu2017, Liu2022a}. The $R-L$ relation, on the other hand, is a natural consequence of photoionization physics, and its intrinsic scatter in such symmetric configurations is often attributed to variations in the spectral energy distribution of the ionizing continuum \citep{Korista2004, Fonseca2020, Shen2024, Wu2025a}.

However, the discovery of double-peaked broad emission-line profiles in a population of AGNs that typically show low accretion rates, $\dot{m}\equiv L_{\rm bol}/L_{\rm Edd}$,\footnote{$L_{\rm bol}$ is the bolometric luminosity and $L_{\rm Edd}$ is the Eddington luminosity.} challenges these simple BLR configurations \citep[e.g.,][]{Storchi1993, Eracleous1994, Shields2000, Wang2005}. Some of these objects exhibit extremely broad lines (with velocity widths exceeding 10,000 km/s) and stronger blueshifted peaks, features that can be well described by high-velocity motions in the outer regions of accretion disks, where Doppler boosting is significant \citep{Chen1989}. In many others, the profiles show stronger red peaks, velocity widths of only a few thousand km/s, or time-variable relative intensities between the blue and red peaks. These complexities have led to the proposal of non-uniform BLR structures, such as spiral arms or hot spots embedded within the disk \citep{Storchi2003}. While these models can reproduce some of the observed variations in the line profiles, the expected periodic variability from bulk motion of such structures has not been detected \citep{Lewis2010}.

The elliptical-disk model provides a compelling alternative interpretation \citep{Eracleous1995}. In this configuration, all gas clouds move in Keplerian orbits confined to a common plane with a fixed orientation, allowing for a long-lived stable BLR. Because the SMBH resides at one focus of the elliptical orbits, gas at the apocenter is located farther from the ionizing source and naturally produces longer time delays. This geometry can explain the asymmetric velocity-resolved lags commonly observed in RM campaigns \citep{Denney2009, Bentz2008, Villafana2022, Feng2021b, Feng2025, Wang2025a}. Long-term stability in both the double-peaked line profiles and velocity-resolved lags observed in some sources \citep[e.g., NGC 3516;][]{Popovic2002, Denney2009, Feng2021b} supports this scenario.

\begin{figure*}[t]
    \includegraphics[scale=0.72]{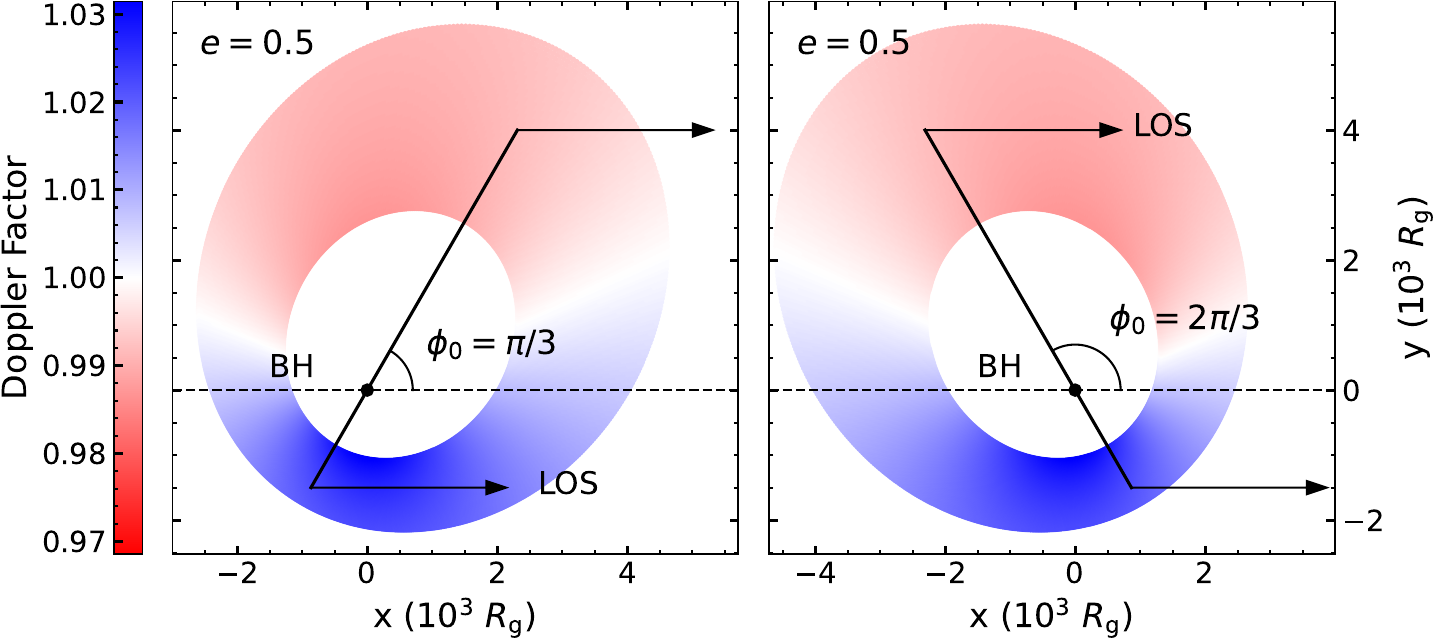}
    \caption{A schematic illustration of elliptical-disk BLR model. In both panels, the observer is located at infinity along the positive $x$-axis. The BLR gas is assumed to rotate counterclockwise. The disk inclination and eccentricity are fixed at $i = \pi/3$ and $e = 0.5$, respectively, and the color indicates the Doppler factor. The left and right panels correspond to disk orientations of $\phi_0 = \pi/3$ and $\phi_0 = 2\pi/3$, respectively. The solid line passing through the black hole illustrates the propagation path of a continuum photon to the pericenter and apocenter, while the arrowed line indicates reprocessed emission-line photons traveling toward the observer. \label{fig:1}}
\end{figure*}

More direct evidence for an elliptical BLR comes from recent ionization mapping studies. \citet{Li2024} reported a significant correlation between velocity-resolved lags and Balmer decrements across the emission-line profile. Since the Balmer decrement is inversely correlated with the incident ionizing photon flux, this trend suggests that redshifted/blueshifted gas with longer lags resides at larger physical distances from the central engine, rather than being a projection effect caused by inflow/outflow motions. Furthermore, BLR eccentricity may lead to a spatial offset between the BLR and the hot dust source, consistent with the GRAVITY interferometric observations of NGC 3783 \citep{Gravity2021}.

With the advent of large spectroscopic surveys, numerous double-peaked AGNs have been identified \citep[e.g.,][]{Eracleous2003, Strateva2003}, and recent studies suggest that such complex line profiles may be ubiquitous among AGNs \citep[e.g.,][]{Storchi2017, Nagoshi2024, Wu2025b}. In this case, even if the gas follows purely virial motion, the peculiar emission-line profiles would significantly affect the estimation of the virial factor \citep{Peterson2004}. Moreover, the RM technique relies on measuring time delays, which are sensitive to the light-travel path \citep{Peterson2014, Mangham2019}. Thus, a non-centrally symmetric elliptical-disk BLR can produce observed lags that vary with azimuthal angle (and inclination), even when the intrinsic BLR size remains constant. As illustrated in Figure \ref{fig:1}, the apocenter lies at a larger line-of-sight (LOS) distance for $\phi_0 = 2\pi/3$ than for $\phi_0 = \pi/3$, so the longer light-travel path could modify the weighted lag. These effects imply that both RM and the $R$--$L$ relation may systematically misestimate the characteristic BLR radius and velocity dispersion.

These considerations motivate us to conduct systematic numerical simulations to quantify how elliptical disk geometries influence the inferred $f$ factor and the BLR size derived from RM and the $R-L$ relation, and ultimately, to assess their impact on SMBH mass determinations. This paper is organized as follows. Section \ref{sec:2} describes the elliptical BLR model. Section \ref{sec:3} presents our numerical experiments over a range of model parameters and the corresponding results. In Section \ref{sec:4}, we discuss the implications for black hole mass estimates. Section \ref{sec:5} summarizes our conclusions.

\section{Methodology}\label{sec:2}

\subsection{BLR Model Description}\label{sec:2.1}
Given that the total mass of the BLR is negligible compared to that of the central SMBH, we ignore the self-gravity of gas clouds in constructing our BLR model. This assumption is widely adopted in BLR dynamical modeling \citep[e.g.,][]{Pancoast2011, Rosborough2024}. Meanwhile, to focus on assessing the impact of BLR geometry on radius estimation and the virial factor, we also neglect the effect of radiation pressure. Under these assumptions, each gas cloud is subject only to the gravitational force of the SMBH and moves along a Keplerian elliptical orbit.

Although many previous BLR models incorporate such orbital dynamics, most treat the orbits of individual clouds as independently distributed with random orientations in space. For a BLR system containing more than $10^6$ clouds \citep{Arav1997}, this randomness would lead to frequent orbital crossings and collisions, making it difficult to maintain a long-term stable structure. Moreover, the superposition of numerous randomly oriented orbits would cause the entire BLR to converge toward a centrally symmetric geometry \citep[e.g.,][]{Pancoast2014, Williams2022}. Therefore, to form a long-lived eccentric BLR, the orbits of gas clouds must exhibit an ordered arrangement.

To this end, we adopt the elliptical disk model proposed by \citet{Eracleous1995}. In this model, all gas clouds are assumed to move within a common plane, sharing a uniform orbital eccentricity $e$ and a common orientation of the major axis $\phi_0$. Here, $\phi_0 = 0^\circ$ corresponds to a configuration where the observer views along the major axis from the apocenter. The SMBH is located at the common focus of all elliptical orbits. With this configuration, specifying the pericenter distance $r_{\rm p}$ and the inclination angle $i$ of the disk relative to the LOS ($0^\circ$ for face-on) fully determines both the distance of any cloud to the center and its LOS-projected velocity.

The radial extent of the BLR is bounded by the inner and outer limits of the pericenter distance, $r_{\rm pi}$ and $r_{\rm po}$, respectively. Because the intensity of the ionizing continuum decreases with increasing distance from the central source, we assume that the gas emissivity $j(r)$ at radius $r$ follows a power-law distribution: $j(r) \propto r^{-q}$. This assumption is consistent with results from CLOUDY simulations \citep{Korista2004}, where the index $q$ is a tunable parameter. The total emission-line profile is then obtained by integrating the emissivity-weighted contributions from all regions across the disk.

In the above model, the observed wavelength of radiation from each cloud is determined by its LOS Keplerian velocity, while the emission strength is governed by the local emissivity. As noted by \citet{Bottorff2000b}, thermal motions and micro-turbulence in the BLR can introduce additional velocity dispersion, which smooth out the sub-structures in the emission line profiles. To account for this effect, we follow the approach of \citet{Eracleous1995} by introducing a constant local broadening parameter $\sigma_{\rm loc}$. Furthermore, gas clouds moving at high velocities near the center under strong gravitational potential experience both gravitational redshift and Doppler boosting effects. To model these effects, we adopt the photon four-momentum formalism of \citet{Chen1989}. In particular, the radial component can be written as
\begin{equation} \label{eq:2}
p^r = s_r \left[ 1 - \frac{b^2}{r^2} \left( 1 - \frac{2\mbh}{r} \right) \right]^{1/2}
\end{equation}
where $b$ is the impact parameter, \mbh\ is the black hole mass, and $r$ is the radial distance from the black hole. Here we introduce an explicit sign factor, $s_r = \pm1$, to encode the photon's initial radial propagation direction along the geodesic. For the `near side' of the disk relative to the observer ($-\pi/2 < \phi < \pi/2$), photons are emitted outward and we thus set $s_r = 1$. For the `far side' ($\pi/2 < \phi < 3\pi/2$), rays reaching the observer are initially directed inward before being bent outward by the black hole potential, and we therefore set $s_r = -1$. This choice enables a accurate treatment of the radial Doppler component in eccentric orbits. This treatment is slightly different from \citet{Eracleous1995}, who adopted $s_r = 1$ for both branches of $p^r$. Figure \ref{fig:1} presents a schematic illustration of the elliptical disk BLR model for visualization.

\begin{deluxetable}{lcc}
\tablecaption{Model Input Parameters and Ranges Explored \label{tab:1}}
\tablewidth{0pt}
\tablehead{
\colhead{Parameter Definition} & \colhead{Fiducial$^{\dagger}$} & \colhead{Ranges}
}
\startdata
Black hole mass ($\log \mbh/M_{\odot}$) & 8.0 & 6.5 -- 9.5 \\
Accretion rate ($\dot{m}$) & - & $10^{-3}$ -- $10^{-1}$ \\
Orbital eccentricity ($e$) & 0.5 & 0 -- 0.9 \\
Major-axis orientation ($\phi_0$) & $\pi/2$ & 0 -- $2\pi$ \\
Inclination angle ($i$) & $30^{\circ}$ & 0 -- $60^{\circ}$ \\
Emissivity power-law index ($q$) & 2.0 & - \\
Inner pericenter distance ($r_{\rm pi}$) & $300 R_{\rm g}$ & - \\
Pericenter radial ratio ($r_{\rm po}/r_{\rm pi}$) & 25 & - \\
Local broadening ($\sigma_{\rm loc}$) & $500 \rm ~km~s^{-1}$ & - \\
\enddata
\tablecomments{Summary of the input parameters used in our elliptical-disk BLR simulations. $^\dagger$ The fiducial parameters are used to determine the ratio between the BLR inner radius and the $R_{\rm BLR}$ derived from the $R$--$L$ relation.}
\end{deluxetable}

\subsection{Model Parameter Setup}\label{sec:2.2}
We treat the eccentricity $e$, the major-axis orientation $\phi_0$, and the inclination angle $i$ as the primary free parameters in our model. The orientation angle $\phi_0$ is assumed to be randomly distributed and is therefore sampled uniformly in the range $[0, 2\pi]$. The eccentricity is allowed to vary from circular orbits ($e = 0$) up to highly elliptical configurations with $e = 0.9$.

For a disk-like BLR, the inclination angle $i$ is one of the most influential yet poorly constrained parameters, as it governs the LOS projection of the velocity field. X-ray observations suggest that the torus covering factor in AGNs is roughly 50\%, implying that Type I AGNs are preferentially viewed at inclinations $i \lesssim 45^\circ$ \citep{Mateos2017}. In addition, polarization measurements indicate that some unobscured AGNs may be observed at inclinations as high as $60^\circ$ \citep{Marin2014}. To ensure a comprehensive exploration of the parameter space, we restrict $i$ to lie below $60^\circ$.

Previous line-profile fitting studies suggest that the emissivity power-law index, $q$, can range from 1.5 to 5 \citep{Eracleous1995}. In this work, we adopt a fixed value of $q = 2$, motivated by the expectation that, for non-overionized gas, the emissivity should follow the radial dilution of ionizing radiation in vacuum, i.e., $r^{-2}$.

\begin{figure}[t]
\includegraphics[scale=0.71]{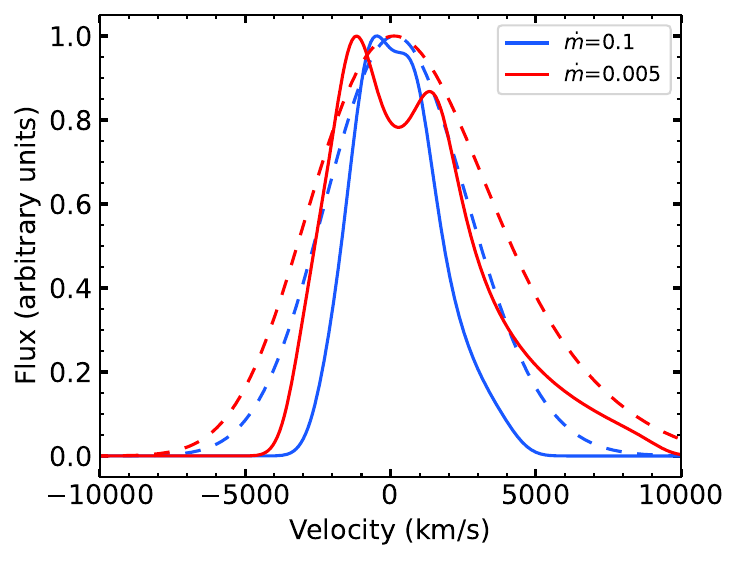}
\caption{Profiles for different local broadening values $\sigma_{\rm loc}$. Solid and dashed lines represent $\sigma_{\rm loc} = 500\,{\rm km\,s^{-1}}$ and $2000\,{\rm km\,s^{-1}}$, respectively. Blue solid line: FWHM = 3540.2 ${\rm km\,s^{-1}}$, $f = 1.1$. Blue dashed line: FWHM = 5750.5 ${\rm km\,s^{-1}}$, $f = 0.4$. Red solid line: FWHM = 5769.5 ${\rm km\,s^{-1}}$, $f = 1.4$. Red dashed line: FWHM = 7381.5 ${\rm km\,s^{-1}}$, $f = 0.8$. $f$ is calculated from the FWHM.
\label{fig:2}}
\end{figure}

The characteristic scale of the BLR is determined based on the empirical $R$--$L$ relation from \citet{Bentz2013}, which is primarily established for low-accretion-rate AGNs:
\begin{equation} \label{eq:3}
R_{\rm BLR} \approx 33.65 \left( \frac{L_{5100}}{10^{44} \, {\rm erg \, s^{-1}}} \right)^{0.53} {\rm lt-days},
\end{equation}
where $L_{5100}$ is the monochromatic luminosity at 5100 \AA. \citet{Richards2006} found an empirical bolometric correction of $L_{\rm bol}\simeq 10\,L_{5100}$. Therefore, for a given black hole mass and accretion rate, $L_{5100}$ and hence $R_{\rm BLR}$ can be directly inferred from Equation (\ref{eq:3}). \citet{Naddaf2020} computed the BLR inner and outer radii in a realistic ``failed radiatively accelerated dusty outflow" model and found that, for low-accretion-rate AGNs, the outer-to-inner radius ratio typically lies in the range 20--30. We therefore adopt a fiducial value of 25 in our calculations.\footnote{Although this ratio can exceed 50 when the accretion rate approaches the Eddington limit ($\dot{m}\sim 1$), our work primarily focuses on sub-Eddington accretion (see Section \ref{sec:4.1}).}

In an eccentric disk geometry, the lag-derived radius depends on multiple geometric parameters. As a result, a given luminosity can correspond to different $R_{\rm BLR}$ values depending on the BLR configuration. To facilitate comparison with the $R$--$L$ relation, we adopt a fiducial configuration with parameters listed in Table \ref{tab:1}: a moderate $e = 0.5$, a representative Type 1 AGN $i = 30^{\circ}$, a $\phi_0 = \pi/2$ such that the pericenter and apocenter are at the same distance from the observer, and $r_{\rm pi} = 300 R_{\rm g}$, similar to those inferred from line-profile fitting \citep{Chen1989}. Here $R_{\rm g} = G\mbh/c^2$ is the gravitational radius. Then, the BLR inner pericenter distance satisfies $r_{\rm pi} = R_{\rm BLR}/12.9$ for any given luminosity.

Spectral fitting of AGN emission lines suggests that the local broadening parameter $\sigma_{\rm loc}$ typically ranges from 500 to 2000 km s$^{-1}$. However, larger values of $\sigma_{\rm loc}$ can smear out substructures in the broad-line profile, potentially biasing the measurement of line widths (see Section \ref{sec:3.1}). To better isolate the influence of disk geometry and inclination on the virial factor, we adopt the minimal value $\sigma_{\rm loc} = 500$ km s$^{-1}$. Table \ref{tab:1} lists all model parameters, together with their fiducial values and the ranges considered in this work. Figure \ref{fig:2} shows a representative example of the simulated emission-line profiles.

\begin{figure}[t!]
\includegraphics[scale=0.73]{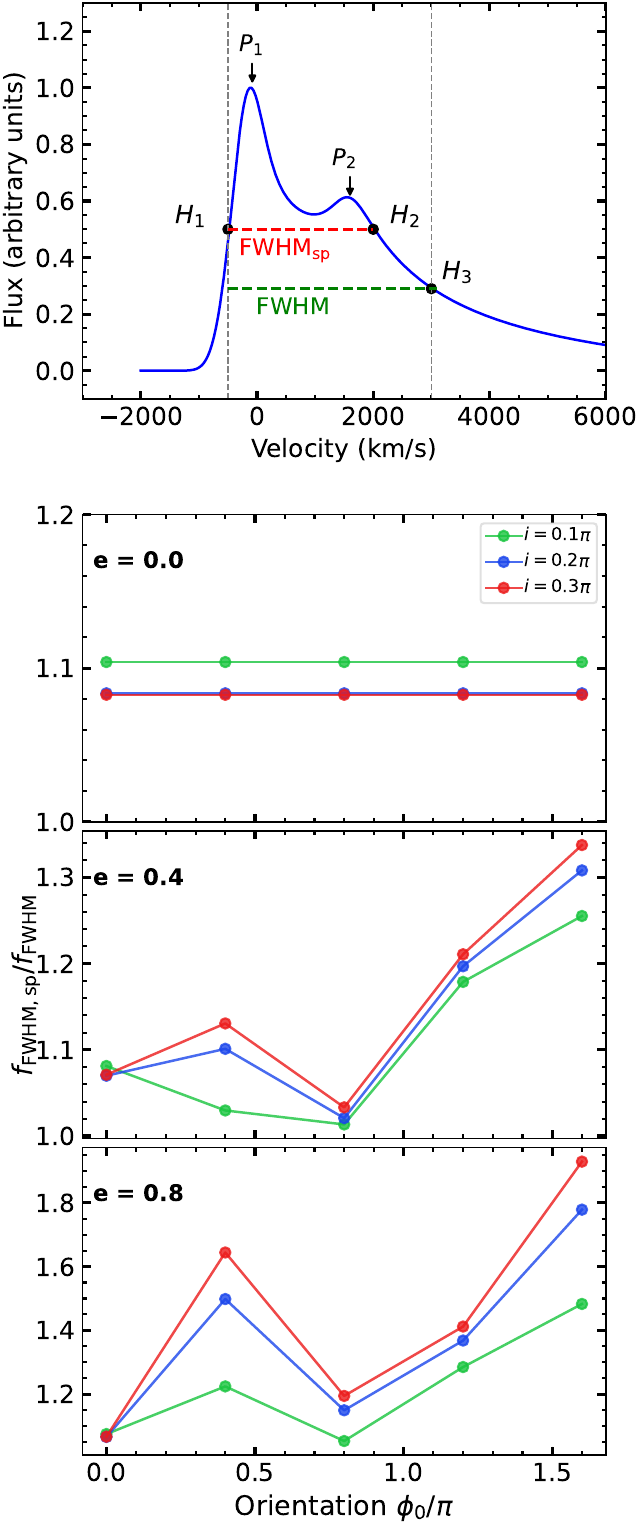}
\caption{In the upper panel, we illustrate the difference between measuring the FWHM using only the single highest peak ($P_1$) and measuring the FWHM by identifying the half-maximum points of both peaks ($P_1$ and $P_2$). The subscript ``sp" denotes ``single peak". In the three lower panels, we present the differences in the virial factor $f_{\rm FWHM}$ resulting from the two methods of computing the FWHM across different parameter sets. 
\label{fig:3}}
\end{figure}

\subsection{Line Width, Lag, and Virial Factor Measurements} \label{sec:2.3}
The bulk velocity of the BLR is typically characterized using the FWHM or the velocity dispersion ($\sigma$) of the broad emission lines. The line dispersion is computed from the second moment of the 
emission-line profile in velocity space:
\begin{equation}\label{eq:4}
\bar v = \frac{\int v\,F(v)\,dv}{\int F(v)\,dv},\qquad
\sigma^2 = \frac{\int (v-\bar v)^2\,F(v)\,dv}{\int F(v)\,dv}, 
\end{equation}
where $F(v)$ is the emission-line flux as a function of the LOS velocity $v$.

In most spectroscopic observations, broad emission lines appear single-peaked, especially at low spectral resolution, which allows for a straightforward measurement of the FWHM. However, the elliptical disk model adopted in this work generally produces double-peaked emission-line profiles under most geometric configurations. Such profiles are often observed in low-accretion-rate AGNs and can complicate standard FWHM measurements, potentially affecting the accuracy of SMBH mass estimates. Following the prescription of \citet{Peterson2004}, we determine the red and blue boundaries of the FWHM by identifying the velocities at which the flux drops to half of the nearest peak.  We adopt this approach for all FWHM measurements in this study. The top panel of Figure \ref{fig:3} schematically illustrates the FWHM definitions based on either a single peak or two peaks. It is worth noting that the resulting virial factor can differ significantly between these two measurement conventions in an elliptical BLR (see the bottom panel of Figure \ref{fig:3}).

Unlike the local lag calculation in \cite{Peterson1993}, we compute the emissivity-weighted time lag integrated over the entire disk surface. In our model, the line emissivity at each location in the BLR is $j(r)$, and the corresponding light-travel-time delay is $\tau(r,\phi)=r(1-\sin i\cos\phi)/c$. The emissivity-weighted lag is therefore
\begin{equation} \label{eq:5}
\tau = \frac{\int\!\!\int j(r) r (1-\sin i\cos \phi) d\phi dr}{c\times\int\!\!\int j(r) d\phi dr}.
\end{equation}
With our adopted power-law emissivity profile $j(r)\propto r^{-q}$, we obtain
\begin{equation} \label{eq:6}
\tau = \frac{\int\!\!\int r^{1-q} (1-\sin i\cos \phi) d\phi dr}{c\times\int\!\!\int r^{-q} d\phi dr}.
\end{equation}
Here, the factor $(1-\sin i\cos\phi)$ encodes the geometric projection effect for an inclined disk (see Figure \ref{fig:1}).

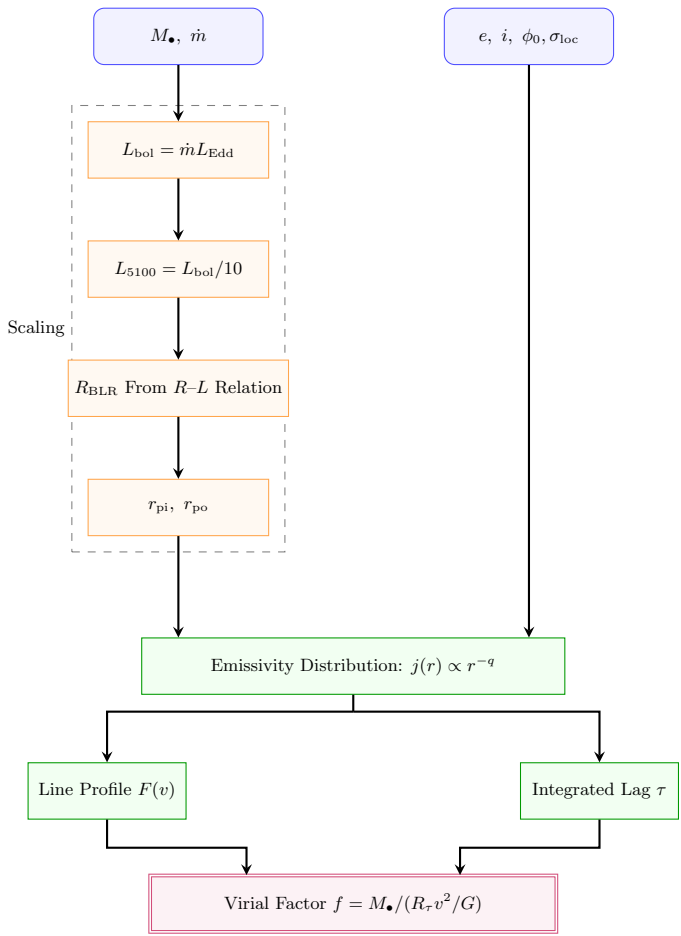
\begin{figure}[ht]
\centering
\begin{tikzpicture}[node distance=1.1cm, font=\small, scale=0.75, transform shape]
\node (in_scaling) [block_input, minimum width=3cm] {$\mbh,~\dot{m}$};
\node (in_geom) [block_input, right=3.2cm of in_scaling, minimum width=3cm] {$e,~i,~\phi_0, \sigma_{\rm loc}$};
\node (p1) [block_scaling, below=1cm of in_scaling, minimum width=3.2cm] {$L_{\rm bol} = \dot{m} L_{\rm Edd}$};
\node (p2) [block_scaling, below=of p1, minimum width=3.2cm] {$L_{5100} = L_{\rm bol} / 10$};
\node (p3) [block_scaling, below=of p2, minimum width=3.2cm] {$R_{\rm BLR}$ From $R$--$L$ Relation};
\node (p4) [block_scaling, below=of p3, minimum width=3.2cm] {$r_{\rm pi},~r_{\rm po}$};
\node (model) [block_sim, below=1.8cm of p4, xshift=3.1cm, minimum width=7.5cm] {Emissivity Distribution: $j(r) \propto r^{-q}$};
\node (spec) [block_sim, below left=1.2cm and -0.8cm of model, minimum width=2.8cm] {Line Profile $F(v)$};
\node (lag) [block_sim, below right=1.2cm and -0.8cm of model, minimum width=2.8cm] {Integrated Lag $\tau$};
\node (f_factor) [block_final, below=3.2cm of model, minimum width=7.2cm] {Virial Factor $f = \mbh / (R_{\tau} v^{2} / G)$};
\draw [arrow] (in_scaling) -- (p1);
\draw [arrow] (p1) -- (p2);
\draw [arrow] (p2) -- (p3);
\draw [arrow] (p3) -- (p4);
\draw [arrow] (p4.south) -- (model.north -| p4.south);
\draw [arrow] (in_geom.south) -- (model.north -| in_geom.south);
\draw [thick] (model.south) -- ++(0,-0.3) coordinate (branch);
\draw [arrow] (branch) -| (spec.north);
\draw [arrow] (branch) -| (lag.north);
\draw [arrow] (spec.south) -- ++(0,-0.5) -| (f_factor.165);
\draw [arrow] (lag.south) -- ++(0,-0.5) -| (f_factor.15);
\begin{scope}[on background layer]
\node[draw, dashed, inner sep=6pt, black!50, label=left:{\small Scaling}] [fit=(p1) (p4)] {};
\end{scope}
\end{tikzpicture}
\caption{Flowchart of the simulation workload.}
\label{fig:4}
\end{figure}

With the measured time lag $\tau$ and line width (either FWHM or $\sigma_{\rm line}$), we compute the virial product according to Equation \ref{eq:1}. Comparing this \mvp\ to the input black hole mass $\mbh$ used in the simulations allows us to infer the corresponding virial factor $f$ = \mbh/\mvp. Figure \ref{fig:4} presents a flowchart summarizing our complete simulation workflow.

\begin{figure*}[ht]
\includegraphics[scale=0.95]{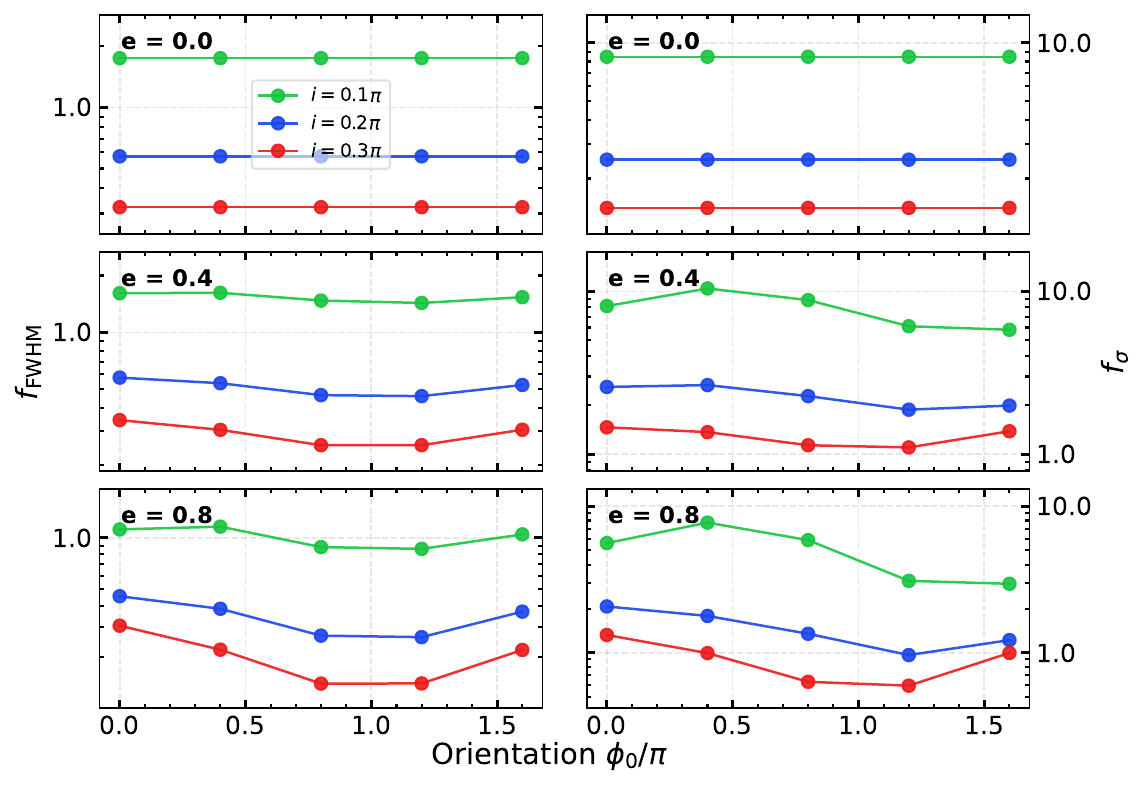}
\caption{The distribution of $f$ under different eccentricities, where $\mbh = 10^8M_\odot$ and $\dot{m} = 0.01$. In the left panel, the virial factor $f$ is calculated based on the FWHM of the emission line. In the right panel, $f$ is calculated based on the line dispersion $\sigma_{\rm line}$. 
\label{fig:5}}
\end{figure*}

\section{Simulation and Results}\label{sec:3}
To investigate how the geometry of an elliptical BLR disk affects the inferred black hole mass, we perform a series of numerical simulations based on the model described in Section \ref{sec:2}. Each simulation corresponds to a specific BLR configuration, characterized by a set of geometric parameters. The goal is to quantify the effects of these parameters on the measured line width, virial factor, and lag.

\subsection{Variations of Virial Factor with Disk Geometry} \label{sec:3.1}
The elliptical disk model naturally leads to more complex emission-line profiles compared to symmetric configurations, potentially introducing biases in the measurement of $f$. To access this effect, we perform a comprehensive set of simulations spanning three inclination angles ($i = 0.1\pi$, $0.2\pi$, $0.3\pi$), three eccentricities ($e = 0$, $0.4$, $0.8$), and five major-axis orientations ($\phi_0 = 0$, $0.4\pi$, $0.8\pi$, $1.2\pi$, $1.6\pi$). The results are shown in Figure \ref{fig:5}.

We find that for circular disks ($e = 0$), the virial factor decreases by a factor of approximately 5 as the inclination increases from $0.1\pi$ to $0.3\pi$. This trend arises from projection effects: at lower inclinations, the LOS velocity is significantly reduced, causing the observed line width to underestimate the true orbital velocity. Consequently, using a constant virial factor leads to an underestimation of the black hole mass. For highly eccentric disks (e.g., $e = 0.8$), the virial factor exhibits substantial variation with disk orientation, even when other parameters are fixed. For example, at $i = 0.2\pi$, the virial factor can vary by a factor of 2 across different major axis orientations $\phi_0$. When combined with inclination effects, the elliptical geometry can induce changes in $f$ by more than an order of magnitude.

\citet{Peterson2004} recommended that the FWHM of double-peaked profiles should be measured using both peaks (as described in Section \ref{sec:2.3}), and further suggested that velocity dispersion measurements offer more precise black hole mass estimates based on empirical virial relations. However, in practice, observational data often come from low-resolution spectra, which tend to smooth out the detailed structure of broad-line profiles, yielding apparently single-peaked lines \citep[e.g.,][]{Peterson2004, Feng2021b, Li2021}. This limitation cannot be fully corrected by simply accounting for instrumental broadening. Moreover, our simulations demonstrate that local broadening, which originates from thermal and turbulent motions of the BLR gas, produces an identical smoothing effect, preventing accurate FWHM measurements even with high-resolution spectroscopy.

To examine these effects, we first analyze how the emission-line profiles evolve with the local broadening parameter. As shown in Figure \ref{fig:2}, increasing $\sigma_{\rm loc}$ transforms double-peaked profiles into single-peaked ones, and increases the observed line width. We then compare the virial factors derived from single-peak FWHM measurements ($f_{\rm FWHM,sp}$) to those obtained using the full double-peaked profiles ($f_{\rm FWHM}$) across different inclinations, eccentricities, and orientations.

The right panel of Figure \ref{fig:3} illustrates a representative case for a SMBH with $\mbh = 10^8 M_\odot$ and an accretion rate of $\dot{m} = 0.01$. In the case of a circular disk, the ratio $f_{\rm FWHM,sp}/f_{\rm FWHM}$ remains constant across different inclinations. In contrast, for highly eccentric disks, significant discrepancies emerge, especially at high inclinations, where the ratio can approach a factor of two. When combined with the line-broadening effect, even moderately eccentric disks can induce deviations in the virial factor by nearly a factor of 3.

Interestingly, when comparing FWHM-based virial factors with those based on velocity dispersion ($f_\sigma$), we find no clear evidence that $\sigma$-based measurements provide more precise results (see also Figure \ref{fig:5}). Both methods exhibit comparable scatter, suggesting that neither approach is inherently superior in the context of elliptical BLR geometries.

\subsection{Effects of BLR Geometry on Lag Measurements} \label{sec:3.2}

Variations in BLR geometry can change the spatial distribution of emissivity, and consequently introduce uncertainties in the measurement of time delays (Figure \ref{fig:1}). To evaluate how projection effects in elliptical disks influence the inferred BLR size, we perform 300 simulations with a fixed eccentricity of $e = 0.5$. The simulations uniformly sample the parameter ranges listed in Table \ref{tab:1}.

\begin{figure*}[t]
\includegraphics[scale=1.0]{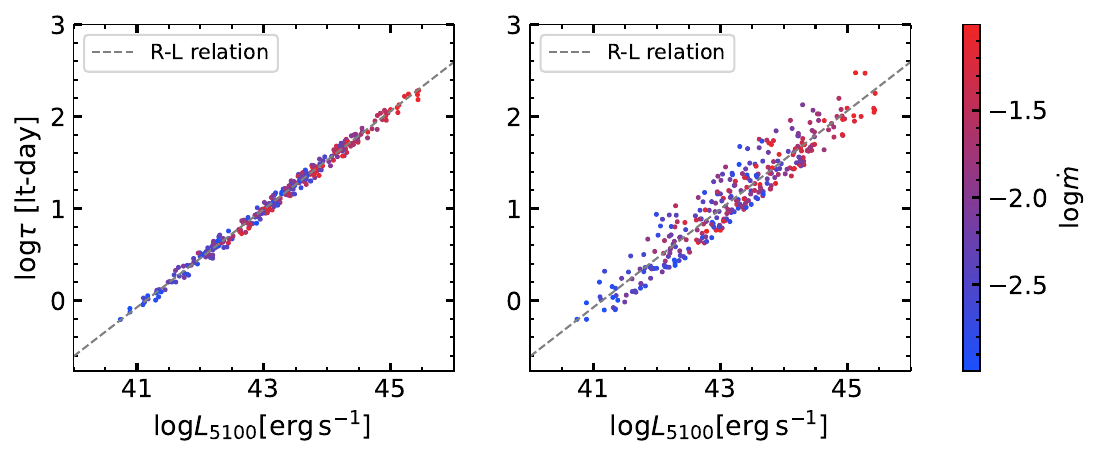}
\caption{The relation between 5100\AA~luminosity and time lag of emission line. The colored dots are given by our calculation. The gray dashed line is the empirical relation from \citet{Bentz2013}. The left panel use a fixed eccentricity $e = 0.5$, while the right panel use the random eccentricities in (0, 0.9).
\label{fig:6}}
\end{figure*}

We compare the lag-derived radii from our simulations with the empirical $R$--$L$ relation from \citet{Bentz2013}. The left panel of Figure~\ref{fig:6} shows that the scatter is remarkably small---only 0.05 dex. This indicates that, for a fixed eccentricity, variations in lag due to changes in inclination and orientation have a relatively minor impact on the inferred BLR size. However, this low scatter may represent a special case. When we allow the eccentricity to vary freely in the range [0, 0.9], the situation changes significantly. As shown in the right panel of Figure \ref{fig:6}, the scatter in the $R$--$L$ relation increases to 0.18 dex, which is similar to that of \citet{Bentz2013}. This enhancement arises because eccentricity fundamentally reshapes the radial distribution of ionized gas, which in turn affects the emissivity-weighted time delays.

While the intrinsic scatter in the $R$--$L$ relation has traditionally been attributed to differences in AGN continuum properties or BLR ionization conditions \citep{Korista2004,Fonseca2020,Shen2024}, our findings demonstrate that geometric diversity can also account for a substantial portion of the observed 0.13--0.19 dex scatter reported in previous studies \citep{Bentz2013}.

\section{Discussion}\label{sec:4}

\subsection{Applicability and Scope of Results} \label{sec:4.1}

Observationally, low-accretion-rate AGNs follow a tight $R$--$L$ relation established by \citet{Bentz2013}, whereas their high-accretion-rate counterparts exhibit systematically shorter lags \citep{Du2018, Li2021}. To address this discrepancy, \citet{Du2019} proposed incorporating the \feii/\hb\ flux ratio as an additional parameter to unify the relation. However, \citet{Feng2025} showed that this correction may inadvertently exacerbate the scatter for low-accretion-rate sources. These conflicting results suggest that low- and high-accretion-rate AGNs may be governed by fundamentally different physical processes in their BLR scale-luminosity relationships, making a single, unified $R$--$L$ relation inadequate.

Although the elliptical-disk BLR model adopted here does not, in principle, require any specific accretion-rate regime, the detection of double-peaked broad emission-line profiles provides an observational constraint on its applicability. These distinctive profiles are predominantly observed in low-accretion-rate AGNs \citep[e.g.,][]{Eracleous2003}. Therefore, the elliptical-disk geometry considered in this work may be most relevant to low-accretion-rate systems.

Furthermore, our simulated mean virial factors are systematically smaller than those calibrated using the black hole mass--stellar velocity dispersion ($M$--$\sigma_{*}$) relation. Specifically, we obtain average values of $f_{\rm FWHM} = 0.71$ and $f_\sigma = 3.43$, which are lower than the values of $f_{\rm FWHM} = 1.1$ and $f_\sigma = 4.5$ reported by \citet{Woo2015}. Despite the substantial intrinsic scatter in both $f$ values, this discrepancy may reflect genuine differences between accretion-rate regimes.

For example, \citet{Mejia2018} used spectral energy distribution (SED) fitting of an accretion disk model and found that broader emission lines correspond to smaller virial factors. Since line width typically decreases with increasing accretion rate \citep[e.g.,][]{Shen2014}, this trend suggests that low-accretion-rate systems naturally have smaller virial factors. Moreover, independent dynamical modeling also yields similarly small virial factors, with mean values of $f_{\rm FWHM} \approx 0.8$ and $f_\sigma \approx 3.8$ \citep[see Table 4 in][]{Wang2026}. These AGNs are generally characterized by low accretion rates and correspondingly small values, e.g., $f_{\rm FWHM} = 0.26$ for NGC 5548 ($\dot{m} \sim 0.01$), $f_{\rm FWHM} = 0.58$ for Arp 151 ($\dot{m} \sim 0.1$), $f_{\rm FWHM} = 0.21$ for NGC 6814 ($\dot{m} \sim 0.03$), and $f_{\rm FWHM} = 0.48$ for SDSS J2222+2745 ($\dot{m} \sim 0.05$) \citep{Pancoast2014, Williams2021}. This quantitative agreement suggests that elliptical-disk BLR geometries may be more common in low-accretion-rate AGNs.

\subsection{Implications for SMBH Mass Measurements}\label{sec:4.2}

Our simulations demonstrate that in asymmetric BLR geometries, the observed time delays can vary significantly due to projection effects, even when the intrinsic BLR size remains unchanged. This finding not only offers new insights into the origin of the intrinsic scatter in the $R$--$L$ relation, but also challenges conventional assumptions regarding the sources of uncertainty in SMBH mass measurements.

In traditional RM frameworks, an axisymmetric BLR geometry is typically assumed, under which the observed lag is believed to reliably trace the physical scale of the BLR. Consequently, the intrinsic scatter in the $R$--$L$ relation has generally been attributed to differences in ionization conditions, while uncertainties in black hole mass estimates are mainly ascribed to inclination effects and non-virial motions.

The diverse kinematic properties of the BLR are often inferred through velocity-resolved lag analysis. However, \citet{Li2024} argued that such diagnostics suffer from degeneracies between BLR kinematics and geometry, making it difficult to identify real inflows or outflows. \citet{Feng2024} further demonstrated that even when velocity-resolved signatures exhibit rapid variations, the BLR can still obey a virial motion. These findings suggest that non-virial motions may not be the dominant source of uncertainty in the virial factor, at least in some AGNs.

\begin{figure}[ht]
\includegraphics[scale=0.61]{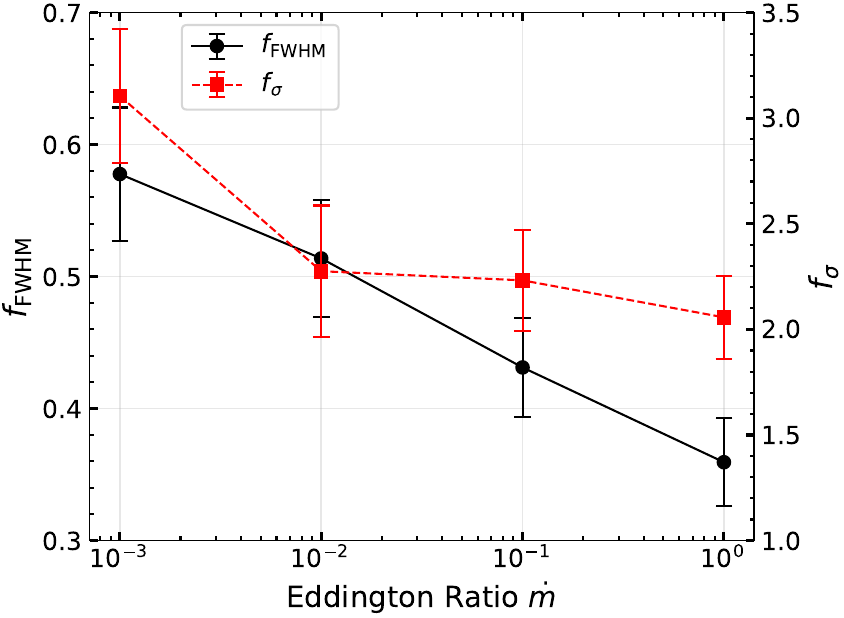}
\caption{The distribution of $f$ under different mass accretion rates $\dot{m}$, where $\mbh = 10^8M_\odot$, $e = 0.4$ and $i = 0.2\pi$. Each data point represents the mean value averaged across different orientations $\phi_0$, and the error bars denote the corresponding standard deviations.
\label{fig:7}}
\end{figure}

Radiation pressure has been widely invoked as a primary driver of virial factor variations. Even in the absence of outflows, it can counteract the gravitational force acting on the BLR gas, thereby reducing its orbital velocity and leading to an underestimation of \mbh\ when a constant virial factor is assumed.  Some statistical studies have found that the virial factor tends to decrease with increasing accretion rate \citep[e.g.,][]{Mejia2018, Caglar2023, Liu2024}, which is commonly interpreted as evidence of radiation pressure. 

Notably, our simulations reveal a similar trend, even within the low-accretion-rate regime. As shown in Figure \ref{fig:2}, higher accretion rates produce narrower emission lines and yield smaller inferred $f$ factors, despite the absence of radiation pressure in our models. The narrowing effect mimics the observational consequences of radiation pressure, but instead arises from BLR photoionization: increasing $\dot{m}$ expands the line-emitting region to larger radii, where orbital velocities are lower, naturally yielding narrower profiles. However, we caution that such photoionization mechanism can also increase the responsivity-weighted BLR size, i.e., the well-known AGN ``breathing'' behavior \citep{Cackett2006, Wang2020}. If breathing correspondingly increases the measured lag approximately as $\tau \propto v^{-2}$, then the increase in $\tau$ can compensate for the decrease in line width, keeping the virial product in Equation \ref{eq:1} nearly constant. In this case, the apparent narrowing of emission lines at higher $\dot{m}$ would not require invoking a systematic adjustment of $f$ to maintain consistent \mbh\ estimates.

In our simulations, the apparent decrease of $f$ with increasing $\dot{m}$ is instead primarily driven by the increasing importance of local broadening $\sigma_{\rm loc}$. At larger BLR radii, the bulk velocity is relatively small, and the contribution from $\sigma_{\rm loc}$ becomes more significant. This additional broadening increases the observed line width and can therefore bias the inferred $f$ factor toward smaller values. To further assess this trend, we examine the dependence of $f$ on $\dot{m}$ in Figure~\ref{fig:7}, which shows a clear anti-correlation between the them.

Taken together, our results fundamentally reshape the understanding of systematic uncertainties in RM-based black hole mass measurements: even in a virialized BLR, its geometry and emissivity structure can systematically bias the inferred BLR size and black hole mass.

\subsection{Model Limitations and Future Work} \label{sec:4.3}

While our simulations provide valuable insights into the geometric effects on SMBH mass estimation, several simplifications in the current framework should be kept in mind.

First, our BLR model neglects the vertical thickness of the disk, assuming that all emitting clouds orbit within a single plane. However, spatially resolved observations from GRAVITY suggest that the BLR is better represented by a disk with finite vertical extent \citep{Gravity2021}. Moreover, observational studies indicate that the BLR kinematics and geometry may vary radially \citep{Feng2025}, and some AGNs may even host multiple distinct BLR components \citep[e.g., a disk-like structure combined with a thicker or more isotropic component;][]{Strateva2003, Nagoshi2024, Wu2025c}. These additional components and vertical structures could complicate the observed line profiles, emissivity-weighted lags, and the inferred virial factors.

Second, to isolate the effects of spatial distribution on RM observables, we adopt purely Keplerian motion throughout this work. In reality, BLR gas dynamics can be influenced by various physical processes, including radiation or magnetic pressure, external perturbations, and cloud-cloud collisions, etc. These mechanisms may introduce deviations from virial motion and give rise to inflows, outflows, or turbulent components, which can affect the line width.

Third, the emissivity distribution in our model is prescribed as a simple power-law function of radius, rather than being derived self-consistently from photoionization calculations. In practice, the ionization structure of the BLR depends sensitively on gas density, column density, metallicity, and the SED of the ionizing continuum \citep{Korista2004, Wu2023}. Photoionization-based emissivity modeling is essential for capturing the detailed coupling between BLR geometry and emission properties.

Looking ahead, incorporating radiative transfer and photoionization modeling, along with more realistic gas distributions that vary both radially and vertically, will enable more accurate predictions of emission-line profiles and reverberation signatures. In parallel, extending our analysis to include two-dimensional transfer functions will allow for a more detailed comparison with velocity-resolved reverberation mapping data, going beyond the global quantities explored in this work.

\section{Conclusion} \label{sec:5}
In this work, we systematically investigate how elliptical disk geometries of the BLR affect SMBH mass measurements via RM and the $R$--$L$ relation. Using numerical simulations of Keplerian elliptical disks with varying eccentricities, orientations, and inclinations, we quantify the impact of BLR geometry on both the virial factor and the inferred BLR size. Our main findings are as follows:

\begin{enumerate}
\item Even in purely virialized BLRs, geometric asymmetries alone can induce variations in the virial factor by more than an order of magnitude. For circular disks, the virial factor decreases by a factor of $\sim$5 as inclination increases from $0.1\pi$ to $0.3\pi$. For eccentric disks, variations in the orientation angle can introduce additional changes up to a factor of 2 at fixed inclination.

\item Local broadening can smooth out double-peaked profiles and increase the observed line width, resulting in systematic biases in $f$ up to a factor of 3. The ratio between virial factors derived from single-peaked versus double-peaked FWHM measurements can reach a factor of 2. Contrary to conventional expectations, we find no consistent advantage in using either FWHM or $\sigma$ for mass estimation in such complex geometries.

\item Asymmetric BLR configurations can systematically change the lag measurements due to projection effects, even with a fixed intrinsic BLR size. These geometric effects can produce a scatter of $\sim$0.18 dex in the $R$--$L$ relation, comparable to the 0.13--0.19 dex intrinsic scatter observed in RM studies.

\item The presence of double-peaked broad-line profiles and the systematically lower $f$ values in our simulations indicate that the elliptical disk model is particularly well suited to low-accretion-rate AGNs.

\item The observed anti-correlation between virial factor and accretion rate can arise from changes in the radial emissivity distribution, rather than being solely driven by radiation pressure.
\end{enumerate}

These findings challenge the conventional view that uncertainties in reverberation-based black hole mass estimates are dominated by non-virial motions or radiation pressure effects. Instead, our results underscore the critical role of BLR geometry in shaping both line profiles and time delays. Future work incorporating photoionization modeling, vertical disk structure, and two-dimensional transfer functions will be essential for refining mass estimates and enhancing our understanding of BLR kinematics and structure.

\begin{acknowledgments}
We thank the anonymous referee for constructive comments that have significantly improved the quality of this work. We also express our sincere gratitude to Xiaolin Yang for his valuable assistance and insightful discussions on the BLR modeling calculations. The work is supported by the National Natural Science Foundation of China (grants 12533005, 12233007, 12573021, 12373018, 12347103) and the science research grants from the China Manned Space Project (No. CMS-CSST-2025-A07). The authors acknowledge Beijing PARATERA Tech CO., Ltd. for providing HPC resources that have contributed to the results reported within this paper. 
\end{acknowledgments}

%% To help institutions obtain information on the effectiveness of their 
%% telescopes the AAS Journals has created a group of keywords for telescope 
%% facilities.
%
%% Following the acknowledgments section, use the following syntax and the
%% \facility{} or \facilities{} macros to list the keywords of facilities used 
%% in the research for the paper.  Each keyword is check against the master 
%% list during copy editing.  Individual instruments can be provided in 
%% parentheses, after the keyword, but they are not verified.

\vspace{5mm}
% \facilities{}

%% Similar to \facility{}, there is the optional \software command to allow 
%% authors a place to specify which programs were used during the creation of 
%% the manuscript. Authors should list each code and include either a
%% citation or url to the code inside ()s when available.

\bibliography{sample631}{}
\bibliographystyle{aasjournal}
\end{document}